\documentclass[
    ,final            
  ]
  {aipproc}

\layoutstyle{6x9}

\begin{document}

\title{Fermion mass and the pressure of dense matter}


\author{Eduardo S. Fraga}{
  address={Instituto de F\'\i sica, 
Universidade Federal do Rio de Janeiro \\
C.P. 68528, Rio de Janeiro, RJ 21941-972, Brazil}
}

\author{Let\'\i cia F. Palhares}{
  address={Instituto de F\'\i sica, 
Universidade Federal do Rio de Janeiro \\
C.P. 68528, Rio de Janeiro, RJ 21941-972, Brazil}
}

\begin{abstract}
We consider a simple toy model to study the effects of finite fermion 
masses on the pressure of cold and dense matter, with possible 
applications in the physics of condensates in the core of neutron stars 
and color superconductivity.
\end{abstract}

\maketitle


The role of finite quark masses in QCD thermodynamics has 
received increasing attention in the last few years. 
In the case of cold and dense QCD, it was generally believed 
that effects of nonzero quark masses on the equation of state 
were of the order of $5\%$, thereby yielding only minor corrections 
to the mass-radius diagram of compact stars \cite{negligible}. 
In fact, mass, as well as color superconductivity gap, contributions to 
the pressure are supressed by two powers of the chemical potential as compared 
to zero-mass interacting quark gas terms. Therefore, assuming a critical 
chemical potential for the chiral transition of the order of a few hundred 
MeV, naively those terms should not matter. However, recent results for the 
thermodynamic potential to two loops have shown that corrections are 
sizable, and may dramatically affect the structure of 
compact stars \cite{Fraga:2004gz}. Moreover, the situation in which 
mass (as well as gap) effects are significant corresponds to 
the critical region for chiral symmetry breakdown in the phase 
diagram of QCD. Hence, not only the value of the critical chemical 
potential will be affected, but also the nature of the chiral transition. 
In particular, if the latter is strongly first-order there might be 
a new class of compact stars, smaller and denser, with a deconfined 
quark matter core \cite{Fraga:2001id}. Of course, contributions due to 
color superconductivity \cite{alford} as well as chiral 
condensation \cite{NJL} will also affect this picture.

In what follows, we study a simple toy model -- cold and dense 
Yukawa theory -- to investigate the influence of fermion masses on the 
pressure. Here, we present a two-loop calculation of the pressure 
with massive fermions in the modified minimal subtraction 
($\overline{MS}$) renormalization scheme \cite{BJP}, and briefly 
comment on possible implications to the physics of condensates 
in the core of neutron stars and effective models for color superconductivity. 
Higher-order corrections and a thorough analysis of renormalization group 
effects will be presented elsewhere \cite{next}.


We consider a gas of massive fermions whose interaction is mediated by a 
real scalar field, $\phi$, with an interaction Lagrangian of the Yukawa form,
$
\mathcal{L}_I= g~\overline\psi\psi\phi \, ,
$
where $g$ is the coupling constant. In the zero-temperature limit, the 
perturbative pressure results in a power series of 
$\alpha_Y\equiv g^2/4\pi$. 
\footnote{Since we are concerned only 
with the zero-temperature limit, there are no odd powers of $g$ coming from  
resummed contributions of the zero Matsubara mode for bosons in the 
perturbative series.} Up to $O(\alpha_Y)$, the first non-trivial 
contributions to the pressure are given by the free massive gas term, 
$P_0$, and the ``exchange diagram'', $P_1$. Using standard methods of
field theory at finite temperature and density \cite{livroKapusta},
one can derive the free gas pressure for fermions of mass $m$, obtaining
in the zero-temperature limit the following form:
\begin{equation}
\lim_{T\to 0}  P_0= 
\frac{1}{12\pi^2} 
\left[ \mu p_f \left(\mu^2-\frac{5}{2}m^2\right)
+\frac{3}{2}
m^4 ~\ln\left(\frac{\mu +p_f}{m}\right) \right] \, , 
\end{equation}
where $\mu$ is the chemical potential and $p_f=\sqrt{\mu^2-m^2}$ 
denotes the Fermi momentum.
The $O(\alpha_Y)$ renormalized correction reads \cite{BJP}:
\begin{equation}
\lim_{T\to 0} P_1=-\frac{\alpha_Y}{4\pi^3}\left[\frac{3}{4}u^2- p_f^4 
+ m^2 \left(3+2 \ln\frac{\Lambda^2}{m^2}\right)u\right]\, ,
\label{p1-final}
\end{equation}
where $u=\mu p_f-m^2 \ln[(\mu+p_f)/m]$ and $\Lambda$ is the 
renormalization scale in the $\overline{MS}$ scheme.


Fig. 1 illustrates the effect of modifying the mass on the total pressure 
to $O(\alpha_Y)$, $P=P_0+P_1$. The choice of range for $\mu$, 
and accordingly for the masses, are inspired by the scales found in the case 
of QCD \cite{Fraga:2004gz}. In the same vein, the coupling is fixed to 
$\alpha_Y=0.3$. It is clear from the figure that mass corrections bring 
significant changes to the pressure, even in the absence of renormalization 
group (RG) running for the coupling and the mass.
The figure also shows the dependence on the renormalization scale $\Lambda$. 
The values chosen are motivated by the ones which appear in QCD, as 
before. Although the effects of varying $\Lambda$ appear to be relatively 
small, it would be premature to conclude that this feature will remain after 
implementing the RG flow. In fact, the results presented in Fig. 2 most 
probably 
underestimate the scale dependence of the full correction, since not only 
the coupling but also the mass will run with $\Lambda$. In the Yukawa 
theory, in contrast to QCD, the effect will become larger as we increase the 
chemical potential. For fixed coupling, larger values of $\Lambda$ 
yield larger modifications in the pressure. However, after the 
inclusion of RG running, this behavior can be mantained, as should be the 
case here, or become 
the opposite, as is the case in QCD, depending on the sign of the 
beta function. Since the $\Lambda$-dependence comes from the term 
$\sim m^2\alpha_Y\ln(\Lambda/m)$ in (\ref{p1-final}), there will be a 
competition between the behavior of the renormalization scale $\Lambda$ 
and that of $m$ and $\alpha_Y$ as functions of $\mu$.
\begin{figure}
\label{fig1}
\includegraphics[height=.25\textheight]{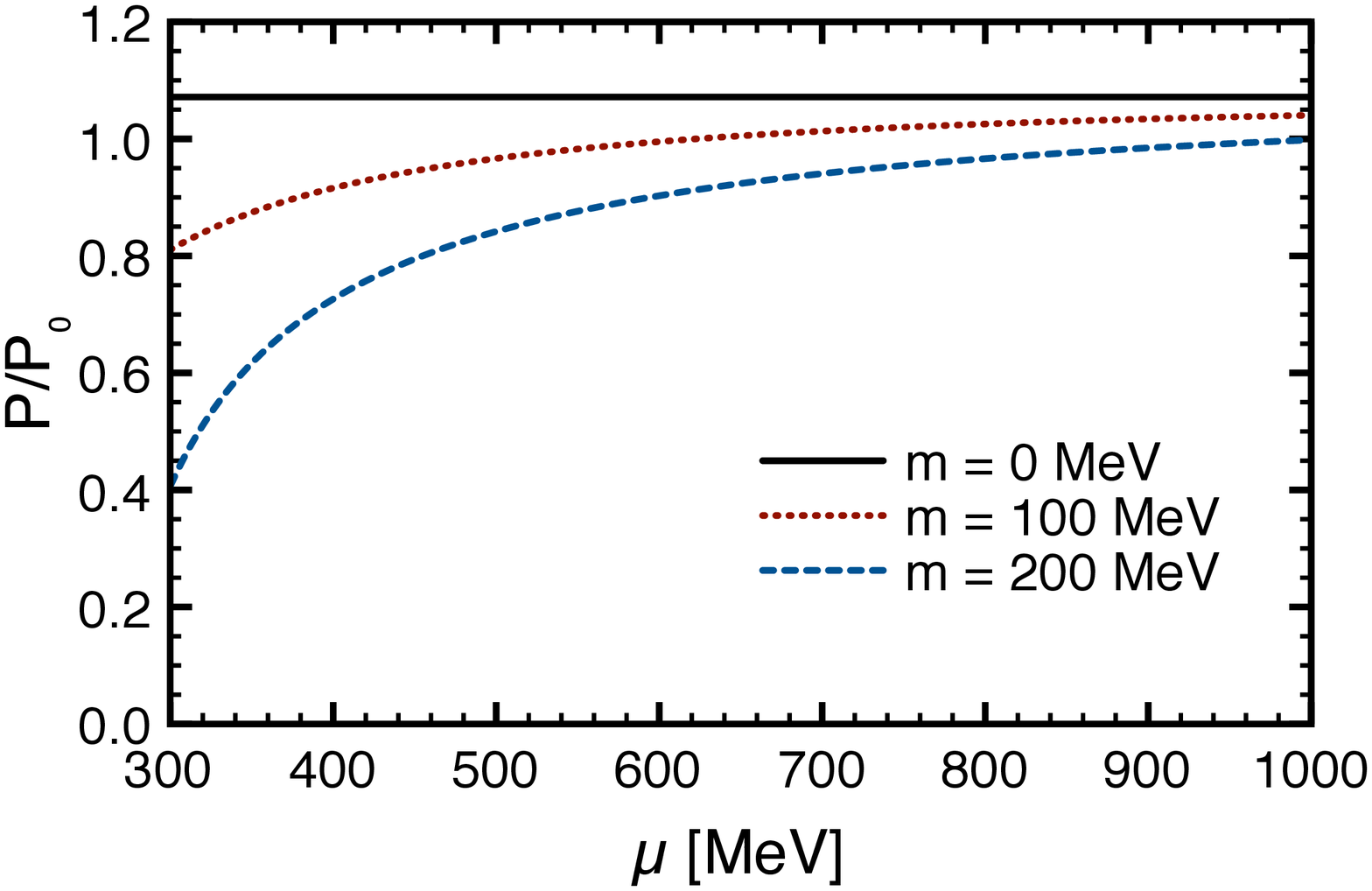}
\includegraphics[height=.25\textheight]{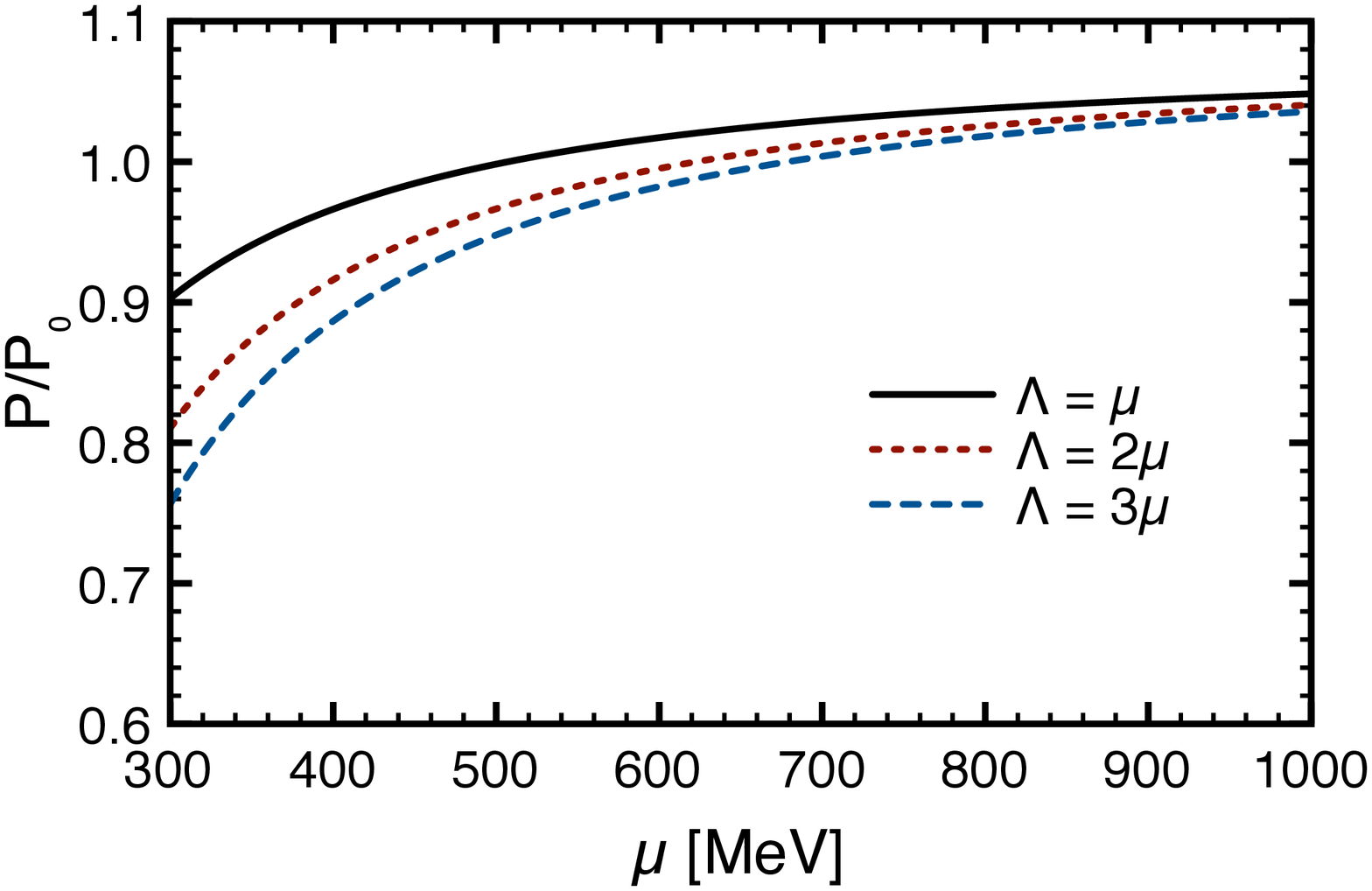}
\caption{Pressure normalized by the free fermion gas pressure as 
a function of the fermion chemical potential. Left: $\Lambda=2\mu$ 
and different values of the fermion mass. Right: $m=100~$MeV and 
different values of the renormalization scale $\Lambda$.}
\end{figure}

Even at two loop order mass effects bring into play logarithmic 
corrections originated in the $\overline{MS}$ subtraction scheme. 
As usual, they bring about a non-physical dependence on the 
renormalization scale $\Lambda$, since one has to cut the perturbative 
series at some order. Higher-order computations in this framework are in 
progress \cite{next}, and will give a better handle on the choice of this 
scale, which in our case should be a function of $\mu$ and $m$. 
On the other hand, one can also choose the scale in a phenomenological 
way in a given model, imposing physical constraints to the equation of 
state, as was done in Ref. \cite{Fraga:2001id} to model the non-ideality 
of QCD at finite density with massless quarks.

The points discussed above might be relevant in the study of effective 
models for the cold and dense matter found in the interior of compact 
stars, especially because the effects seem to be significant near the 
critical region. In the context of the NJL model, e.g., 
it was shown that a self-consistent treatment of quark masses 
strongly affects the competition between different phases \cite{NJL}. 
And the mechanism of pairing in color superconductivity will certainly be 
influenced \cite{Rajagopal:2005dg} by the 
running of nonzero quark masses. The investigation of these issues, as 
well as the effect of nonzero fermion masses in the formation of other 
condensates in neutron star matter, is under way \cite{next}.


\begin{theacknowledgments}
We thank R. D. Pisarski, J. Schaffner-Bielich and C. Villavicencio for 
fruitful discussions.
This work was partially supported by CAPES, CNPq, FAPERJ and FUJB/UFRJ.
\end{theacknowledgments}



\bibliographystyle{aipproc}   


\end{document}